\documentclass[aps,showpacs,twocolumn,floatfix,amsmath,preprintnumbers,nofootinbib,secnumarabic,letter]{revtex4}

\usepackage{epsfig,verbatim,hyperref}
\usepackage{amssymb,amsfonts}
\usepackage{graphicx}

\usepackage[caption=false]{subfig}
\usepackage{color}

\def\jrn#1#2#3#4#5#6{#3 \textbf{#4}, #5 (#6)}  %aip
\def\scn#1#2{\textbf{#1}. --} %epl style
\def\eq{eq.\,}  %epl style

\def\bfl{\begin{flushleft}}
\def\efl{\end{flushleft}}
\def\bfr{\begin{flushright}}
\def\efr{\end{flushright}}
\def\bc{\begin{center}}
\def\ec{\end{center}}
\def\be{\begin{equation}}
\def\ee{\end{equation}}
\def\bse{\begin{subequations}}
\def\ese{\end{subequations}}
\def\ba{\begin{eqnarray}}
\def\ea{\end{eqnarray}}
\def\baa#1{\begin{array}{#1}}
\def\eaa{\end{array}}
\def\bw{\begin{widetext}}
\def\ew{\end{widetext}}

\def\lb#1{\label{#1}}
\def\bit{\begin{itemize}}
\def\eit{\end{itemize}}
\def\bco{\begin{comment}} \def\eco{\end{comment}}
\def\bcs{\begin{cases}}
\def\ecs{\end{cases}}

\def\schrod{Schr\"odinger}

\def\lan{{\cal L}}
\def\lanp{{\cal V}}

\def\Der#1#2{\frac{\drm #1}{\drm #2}}
\def\pDer#1#2{\frac{\partial #1}{\partial #2}}

\def\vena{\boldsymbol{\nabla}}

\def\nc0{\tilde b_0}

\def\enin{{\cal U}}
\def\icip{\mathbb{U}}
\def\U{{\rm U}}
\def\H{{\rm H}}
\def\T{{\rm T}}
\def\cf{{\cal D}}
\def\ent{{\cal S}}
\def\vol{{\cal V}}
\def\vol{V}
\def\ello{{\ell_0}}

\def\drm{d}
\def\dvol{\drm\vol}
\def\dm{{\bar d}}

\begin{document}

\preprint{\small Europhys. Lett. (EPL) \textbf{122}, 39001 (2018)   
%\ \ [DOI: 10.1209/0295-5075/122/39001]
\quad 
[\href{https://doi.org/10.1209/0295-5075/122/39001}{DOI: 10.1209/0295-5075/122/39001}]
}
% [arXiv:17xx.xxxxx]

\title{
Nonlinear wave-mechanical effects in Korteweg fluid magma transport
%Symmetry breaking, fluid fragmentation and foam formation
}

\author{Konstantin G. Zloshchastiev}
%\email{k.g.zloschastiev@gmail.com}
\email{http://bit.do/kgz}
\affiliation{Institute of Systems Science, Durban University of Technology, P.O. Box 1334, Durban 4000, South Africa}

%\email{k.g.zloschastiev@gmail.com, kostiantynz@dut.ac.za}

\begin{abstract} 
Statistical mechanics arguments and Madelung hydrodynamical presentation 
are applied 
to the transport of magma in volcanic conduits.
An effective wave equation with logarithmic nonlinearity
becomes apparent in systems of this kind,
which describes
the flow of a two-phase barotropic Korteweg fluid with capillarity,
and allows multiple eigensolutions  
thus leading to wave-mechanical effects. 
We study spontaneous symmetry breaking in the erupting lava
which flows up the conduit, so that
fluid fragmentation and nucleation of density inhomogeneities occur;
therefore, changing temperature 
can trigger a transition between the ``magma-dissolved gas'' fluid 
and magmatic foam phases.
This phase structure is studied by both analytical and numerical methods.
In the fluid phase, cell-like inhomogeneities occur which are 
described by solitary wave solutions with a Gaussian density profile;
we derive 
the many-body interaction potential for these inhomogeneities.
For the foam phase, we demonstrate existence of topological kink solitons which  
describe bubble-type inhomogeneities; 
their stability is ensured by the conservation of a topological charge.
\end{abstract}

\date{6 Feb 2018 [APS], 20 Feb 2018 [EPL]}%, xx Sep 2017 [arXiv]}
%\date\today

\pacs{91.40.Ft, 47.55.nb, 47.35.Fg
%91.40.-k	Volcanology 91.40.Ft – Eruption mechanisms 47.55.nb	Capillary and thermocapillary flows 47.10.-g	General theory in fluid dynamics 47.20.-k – Flow instabilities 47.35.Fg – Solitary waves
%\\ \textbf{Keywords}: logarithmic BEC, quantum liquid, stability, collective oscillations
}

\maketitle

%\scn{Introduction}{s:in}

An active volcano can be modeled as a hydrodynamic
system consisting of a magma reservoir or chamber which is connected
to a long conduit.
Before eruption, magma is accumulated inside pockets under the Earth's crust. 
Changes in pressure trigger various chemical and physical processes,
such as cavitation and gas diffusion,
which fragment magma, form a foam, eventually leading
to breaking of the crust, followed by a volcanic eruption. 
The ascent of magma is usually described as a
steady one-dimensional flow
of an isothermal compressible fluid in a porous
medium driven by pore compaction and dilation \cite{fo84,mc84,ss84,ssw85,toc90}.

The use of barotropic Korteweg capillary
fluid as a model of magma transport in a conduit was recently proposed in works \cite{dmf03,gl08}.
In this class of theories, commonly referred as diffuse interface models,
the capillary interface is viewed as a diffusion transition
domain of rapid smooth variation of density, while surface
tension is intrinsically incorporated \cite{ds85,dk93,amw98}.
This allows us to describe flows with spontaneous
nucleation, coalescence and breakdown of density inhomogeneities in
two-phase fluids \cite{an96}.\\

\scn{Wave equation}{s:we}
We consider a magmatic fluid flowing in a cylindrical channel,
which is in thermal contact with a reservoir of infinitely large heat capacity so as
to maintain 
constant temperature.
We assume also that
this fluid is formed by partial melting of solids with a mostly low thermal conductivity, such as silica.
Therefore, its microscopic structure can be regarded as
a many-body system of particles, atoms and molecules, whose average potential energy is larger
than kinetic. 
%% (\textit{i.e.}, it is condensate-like).
Then the probability $P$
of a microstate in this system is given by a Boltzmann-type rule,
in which kinetic energy can be neglected compared to potential:
\be\lb{e:statph}
P \propto \exp{(-{\cal E}/T)} \approx \exp{\left(-\U /T\right)}
,
\ee 
where $T$, ${\cal E}$ and $\U$ are, respectively, the temperature, energy and potential energy of a many-body system; here we work in units where the Boltzmann constant $k_B = 1$.

At a microscopic level, this many-body system is described
by a rather large set of particles' evolution equations, 
%therefore certain simplifications are in order here.
but employing the fluid approximation used in continuum mechanics,
we can make a transition to a single equation for the fluid wavefunction
in a Madelung form:
\be\lb{e:fwf}
\Psi = \sqrt\rho \exp{(i S)}
,
\ee
where $\rho = \rho(\textbf{x},t)$ is a fluid density, and
$S = S(\textbf{x},t)$ is a phase which is related to fluid velocity via the gradient: $\textbf{u} \sim \vena S$
(we assume irrotational flow).
%and $\beta$ is a scaling parameter whose meaning to be clarified below.
This wavefunction should not be confused with particle wavefunctions in
quantum mechanics, 
but rather it is a complex-valued function which stores macroscopic information
about the underlying many-body system, such as the density and velocity of a fluid element or parcel
%, as discussed in Ref. 
\cite{ry99}
(another wave-mechanical analogy can be found in, \textit{e.g.}, classical electromagnetic wave theory \cite{sybook09,z16}).
Nevertheless, some mathematical similarities between these categories of functions do exist and will be used
here.
Specifically, the function $\Psi$ obeys a normalization condition
\be\lb{e:norm}
\int_\vol |\Psi|^2 \dvol  = 
\int_\vol \rho\, \dvol = M
%> 0
,
\ee 
where $M$ and $\vol$ are the total mass and volume of the fluid.
If we consider that this function is also a solution of a wave equation,
%containing spatial derivatives of the order at least two,
this poses restrictions,
% upon fluid wave functions, 
which are
somewhat
similar to those of a quantum-mechanical case:
a set of all normalizable fluid wave functions must
constitute a Hilbert space, such as $L^2 (\mathbb{R}^{\dm})$
where $\dm$ is the number of spatial dimensions of the fluid.

To derive an equation for $\Psi$, 
%it is natural to
we 
correlate the probability sample space of an underlying microscopic
many-body system with the configuration space of fluid's degrees of freedom, \textit{i.e.}, one can 
assume a correspondence between probability \eqref{e:statph} and 
fluid density:
$P \sim \rho$
%and take into account this statistical effect upon an energy of system's collective degrees of freedom. 
hence $|\Psi|^2 \sim \exp{\left(-\U /T\right)}$.
From the latter formula, a general expression follows
for an operator of
the potential $\U$:
$%\be
\hat \U 
%\sim - T \ln{(A |\Psi|^2)} 
= 
- (T-T_0) \ln{(\rho/\rho_0)},
$ %\ee
where 
%$A = 1/\rho_0$ is some dimensionful constant, and 
$\rho_0$ and $T_0$ are reference values of, respectively, fluid density and temperature.
%%(one can see that at these values the potential $\U$ changes its sign).

Assuming that our flow is Hamiltonian,
we consider an operator form for its total energy,
which can be written as an identity:
$\hat \H  = \hat \T + \hat \U $,
where $\hat \H = i \eta\, \partial_{t}$ is a Liouville-type generator of time translations,
$\eta$ being a scaling dimensionful parameter (we can assume it positive for definiteness),
$\hat \T \propto \frac{1}{2} \hat{\textbf{p}} \cdot \hat{\textbf{p}}$ is a kinetic energy operator,
$\hat{\textbf{p}} \propto - i \vena$ is a generator of spatial translations,
and $\hat \U$ is a potential energy operator derived above.
These operators act in the above-mentioned Hilbert space of states described by rays $|\Psi \rangle$,
therefore, this identity can be recast in the form
$\hat \H |\Psi \rangle = (\hat \T + \hat \U) |\Psi \rangle$,
which brings us to
the logarithmic wave equation of a \schrod~ type (where the role of a Planck constant is played by $\eta$):
%(we assume time and position given in suitably chosen units):
\ba
i \partial_t \Psi
=
\left[-\frac{\cf}{2} \vena^2
- b 
 \ln\left(|\Psi|^{2}/\rho_0\right)
\right]\Psi
,\label{e:o}
\ea
where $b = (T-T_0)/\eta$ and $\cf$ are real constants.
The latter is a material parameter of the theory
related to surface tension; in wave-mechanical context it acts as
a measure of the inertia of a capillary flow.
It becomes apparent that $T_0$ is the critical
temperature at which the coupling $b$ changes its sign,
so it is related to the phase transition discussed below.

The wave equation must be supplemented with the normalization condition \eqref{e:norm}.
Besides, if the fluid flows inside some kind of vessel or external potential
$V_\text{ext} (\textbf{x},t)$, such as
Earth's gravity,
then a corresponding term $\eta^{-1} V_\text{ext} (\textbf{x},t)\, \Psi$ must be added
to a right-hand side of \eq \eqref{e:o}.
Since we are assuming a leading-order approximation here, 
we can neglect geometrical constraint's effects and consider a trapless flow, \textit{i.e.}, $V_\text{ext} \equiv 0$ for now.
%mentioned above.

%capillary's diffusivity and 

One can easily verify that by substituting \eq \eqref{e:fwf} into \eqref{e:o} 
one recovers 
hydrodynamic laws for mass and momentum conservation for a
two-phase compressible inviscid fluid with internal capillarity whose flow is
irrotational and isothermal \cite{dmf03,gl08}:
\ba
&&
\partial_t\rho 
+ \vena\cdot(\rho \textbf{u})
= 0
,\lb{e:floma}\\&&
\partial_t \textbf{u}
+
\textbf{u} \cdot\vena \textbf{u}
-
\frac{1}{\rho} \vena\cdot \mathbb{T}
=0
,
\lb{e:flomo}
\ea
with $\textbf{u} = \cf \vena S$ and the stress tensor $\mathbb{T}$ of the Korteweg form 
with capillarity \cite{ds85},
which
is used to model fluid mixtures with phase changes
and diffuse interfaces \cite{amw98,an96}:
\be\lb{e:stko}
\mathbb{T} 
=
-\frac{\cf^2}{4 \rho} \vena\rho \otimes \vena\rho 
- \tilde p \, %(\rho)
\mathbb{I}
%=-\frac{\nu}{4 \rho} \vena\rho \otimes \vena\rho 
%+ \left[p (\rho) + \frac{\nu}{4} \vena^2\rho \right] \mathbb{I}
,
\ee
where $\mathbb{I}$ is the identity matrix, 
$%\be
\tilde p %(\rho) 
=
p (\rho) - \frac{1}{4} \cf^2 \vena^2\rho 
%= b \cf \rho - \frac{1}{4} \cf^2 \vena^2\rho 
$ %\ee
is capillary pressure, 
and $p (\rho) =  - %epl is correct
\cf b \rho$ is a barotropic equation of state for fluid 
pressure $p$.
%; we have assumed also that $\textbf{u} = \cf \vena S$. 
The stress tensor is also related to the chemical potential
$\mu$
through the formula:
\be\lb{e:chemp}
\vena\mu = 
- \frac{\eta}{\cf \rho} 
\vena\cdot \mathbb{T} =
- \eta
\vena
\left[
\frac{\cf}{2}\frac{\vena^2\sqrt\rho}{\sqrt\rho}
+
b 
\ln{
%(\rho/\rho_0)
\left(\frac{\rho}{\rho_0}\right)
}
\right]
,
\ee
%where a scale factor $\eta$ is incorporated into $\mu$ for brevity. \eq \eqref{e:chemp} also 
which
indicates that the chemical potential 
can be directly determined from \eq \eqref{e:o}
using a stationary state ansatz $ \Psi (\textbf{x}, t) = \exp{\left(- i \mu t /\eta\right)} \Psi (\textbf{x})$.

By using a standard averaging procedure with respect to the inner product
in the Hilbert space of the fluid wavefunctions $\Psi$, one can show that an averaged
form of \eq \eqref{e:o} can be written
as a formula for a wave-mechanical internal energy of the fluid:
\be\lb{e:freee}
\enin =
\langle  \hat{\H}  \rangle
=
{\cal F} +  T \ent  
,\ee
where
${\cal F} = \langle  \hat{\T}  \rangle 
=
-\frac{\cf}{2 M} \int_\vol \Psi^* \vena^2 \Psi \dvol 
$ 
is a wave-mechanical free energy, the
temperature is counted with respect to a reference value $T_0$,
and 
$%\be\lb{e:shent}
\ent 
%= -\int_\vol |\Psi|^2 \ln{(A |\Psi|^2)} \dvol
=
- \frac{1}{M} \int_\vol \rho \ln{\!(\rho/\rho_0)} \, \dvol
$ %,\ee
is an entropy per mass;
according to aforesaid $\ent \sim
-\int_\vol P \ln P \, \dvol$.
Thus, $T$ and $\ent$ must be thermodynamically conjugated:
\be
T \sim b \sim \left(\pDer{\enin}{\ent}\right)_\vol
,\ee if one regards $\enin$ as a thermodynamical potential.

In summary, \eq \eqref{e:o} is a concise form of writing hydrodynamic equations for Korteweg materials,
which makes it a useful model in studies of the Korteweg-type magmas, especially
considering the substantial amount of information accumulated so far about the properties of 
wave equations with logarithmic nonlinearity in the different branches of 
physics, see works \cite{ros68,bb76,z10gc,dz11,dmz15,szm16,zrz17} and references therein,
including the theory of logarithmic Bose-Einstein condensates \cite{gg15,az11,z12eb,bo15,z17zna}.\\

\begin{figure}[htbt]
\begin{center}\epsfig{figure=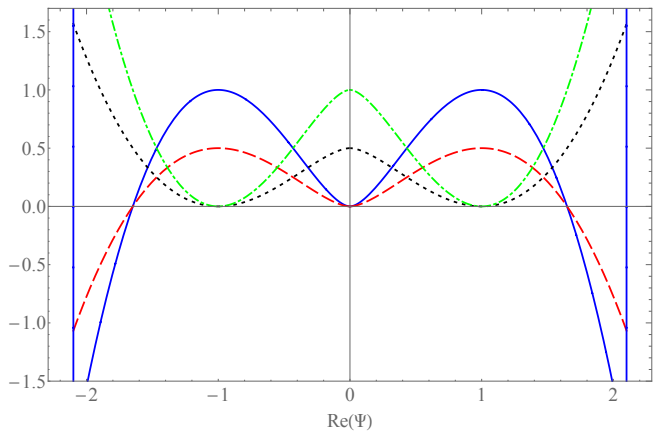,width=  1.0\columnwidth}\end{center}
\caption{
Potential density $\lanp (|\Psi|^2)$ in units of $\rho_0$, versus 
$\text{Re}(\Psi)$ in units of  $|\Psi_e | = \sqrt{\rho_0}$, for the following values
of $b$ (in units of inverse time): $1$ (solid curve), $1/2$ (dashed), $-1/2$ (dotted), $-1$ (dash-dotted).
Two vertical lines represent a condition $|\Psi| \leqslant |\Psi_\text{cut}| < \infty$ which occurs due to the 
%fluid wavefunction's normalization 
constraint \eqref{e:norm}.
%, where a finite cut-off $|\Psi_\text{cut}| > |\Psi_e |$.
%a well with infinite walls, so that the forbidden region starts at $|\Psi| \geqslant |\Psi_\text{cut}| > |\Psi_e |$, where .  
}
\label{f:fpotgau}
\end{figure}

\scn{Symmetry breaking and phase transitions}{s:sb}
Using a variational approach, \eq \eqref{e:o} can be derived
as an Euler-Lagrange equation 
%minimizing the energy functional \eqref{e:freee}:
%following action functional:
for Galilean-invariant Lagrangian density:
\be\lb{e:ftlan}
\lan = 
\frac{i}{2}(\Psi \partial_t\Psi^* - \Psi^*\partial_t\Psi)+
\frac{\cf}{2}
|\vena \Psi|^2
%+ V_\text{ext} |\Psi|^2
+
\lanp (|\Psi|^2)
,
\ee
with potential density given by
\be\lb{e:ftpot}
\lanp (\rho) = -
b \rho
\left[
\ln{(\rho/\rho_0)} -1
\right]
+ \lanp_0
,
\ee
where $\lanp_0 = \lanp (0) $ is a constant whose value can be chosen to ensure
that the potential
equals zero at the local minima;
in our case we can set $\lanp_0 =  \rho_0 (|b|-b)/2$.
For complex values of $\Psi$,
this potential has nontrivial local extrema at $|\Psi_e | = \sqrt{\rho_0}$,
which can be minima or maxima depending on a sign of $b$,
%which changes when temperature goes across a value $T_0$,
cf. Fig. \ref{f:fpotgau}.
This indicates a possibility of symmetry breaking (or restoration),
and the existence of at least two phases in the fluid which correspond
to the positive and negative values of $b$.

It should be noted that for the same values of $b$ and $\rho_0$,
and for the same set of 
boundary conditions, \eq \eqref{e:o}
%, hence \eqs \eqref{e:floma}-\eqref{e:stko}, 
allows multiple normalized eigensolutions.
This can be observed by studying stationary solutions,
because propagating solutions 
%of \eq \eqref{e:o}, which 
can be always generated 
%from stationary ones 
by means of the Galilean transformation 
$\textbf{x} \to \textbf{x} - \textbf{u}\, t$.
The equation for a stationary fluid wavefunction,
$ \Psi (\textbf{x}, t) = 
\exp{\left(- i \omega t \right)} \Psi (\textbf{x})$,
indicates the existence of multiple states in the Hilbert space of the system,
which correspond to different eigenvalues of the wave frequency $\omega = \mu/\eta$.
%Since \eq \eqref{e:o} is \eqs \eqref{e:floma}-\eqref{e:stko},
As discussed after \eq \eqref{e:fwf},
this means that the Korteweg-type magmatic fluid spontaneously selects one of the possible 
eigensolutions,
%(with a certain probability),
which makes it somewhat similar to quantum liquids.
However, the ground state is still the preferred one, as it corresponds to a
minimum of $\omega$ which is also a minimum of
chemical potential.
% (assuming that $\eta$ is positive). 

These aspects will be further discussed in the next two sections,
where we consider the phase structure of the model \eqref{e:o} in details.\\

\scn{Cellular phase}{s:phg}
If temperature of Korteweg magma fluid satisfies the condition
% $(T-T_0)/\eta > 0$ (or 
$T > T_0$
% assuming $\eta$ positive), 
then the nonlinear coupling
$b$ is positive, hence
the potential density $\lanp$ has an upside-down Mexican-hat shape,
with local degenerate maxima at $|\Psi_e | = \sqrt{\rho_0}$, cf. solid and dashed curves in Fig. \ref{f:fpotgau}.

In this case, one solitary wave solution of \eq \eqref{e:o} can be found analytically.
It corresponds to the ground state (\textit{i.e.}, the one with a lowest eigenvalue of 
the frequency $\omega$)
and can be
written
in the form 
of a Gaussian parcel modulated by a plane wave:
\be
\Psi_{(g)} (\textbf{x}, t) = 
\pm 
\sqrt{\rho_g (\textbf{x})}
\exp{\left(- i \omega t + i \textbf{k}\cdot \textbf{x}\right)} 
,
\ee
where 
$\textbf{k}$ is a constant vector,
and 
%$\Psi (\textbf{x})$ is a normalized stationary solution for the ground state reads:
the density's normalized eigenfunction and frequency eigenvalue 
are, respectively:
\ba
\rho_g (\textbf{x})
%= \sqrt{\rho_g (\textbf{x})}
&=&
\tilde\rho
\exp{\left[- \frac{(\textbf{x}- \textbf{x}_0)^2}{\ell^2} \right]}
,\lb{e:gauss}\\
\omega_g
&=&
\mu_g/\eta
=
\frac{1}{2}
\cf \textbf{k}^2
+
b 
\left[
\dm - 
\ln{
 \left(
        \frac{\tilde\rho}{\rho_0}
 \right)}
\right]
,
\ea
where
$\tilde\rho = M/\tilde\vol$,
$
\tilde\vol= 
%[\pi\cf/(2 b)]^{d/2} = 
\pi^{\dm/2} \ell^\dm
% = (\sqrt\pi \ell)^d
$
and
$\ell = \sqrt{|\cf/(2 b)|}$
are the 
density peak value, effective volume and Gaussian width, respectively;
$\textbf{k} = 0$ if $b \not=0$.
%The eigenvalue of $\mu = \mu_{(g)}$ corresponding to the solution \eqref{e:gauss} can be written as: and it must be less than $\mu_{(c)} = b \ln{(\rho_0/\bar\rho)}$, where $\bar\rho = M/\vol $ is an average density of the fluid, otherwise a ground state with respect to $\mu$ would be described not by a Gaussian-shaped wave but by the homogeneous normalized solution  $\Psi_{(c)} (\textbf{x}, t) = \sqrt{M / \vol} \exp{\left(- i \mu_{(c)} t\right)}$.
%Since $\mu_{(c)}$ decreases with decreasing $\bar\rho$, this explains why \eqs \eqref{e:o} and \eqref{e:gauss} are  more useful for describing dense fluids and condensates \cite{gg15,az11,z12eb}.
%In a theory of quantum liquids and Bose-Einstein condensates, solutions \eqref{e:gauss} are usually referred as bright solitons.

Therefore, in this phase, Korteweg-type magma tends to
fragment into clusters of density inhomogeneities 
with a Gaussian profile, referred here as cells.
One can effectively treat these cells 
as classical point particles,
% of mass $m_c$, 
while encoding their nonzero
size in the two-body interaction potential 
$\icip \left( |\textbf{x} - \textbf{x}' | \right)$.
The latter can be approximately derived in the following way.

A spherical
cell of a size $R \sim \ell$ stores an amount
of internal bulk mass-energy 
$
\epsilon_g (R) \propto 
%4 \pi 
\int_0^R \rho_g (r') r'^2 \drm r' 
$.
By direct computation, we obtain
\be
\epsilon_g (R)
\propto
\frac{1}{\ell} (R - \ello) \exp{\left[-(R/\ell)^2\right]}
\left[
1 + {\cal O} \left(R - \ell\right)
\right],
\ee
where
$\ello = \ell
\left[
1/2 + 1/(\text{e} \sqrt{\pi}\, \text{erf}(1))
\right] \approx 0.75 \,\ell$.
%refers to the  point
%is where the dominant term of $\epsilon_g$ changes sign. 
Because each cell tries to maintain its size and mass
when interacting with its environment,
%in order 
to alter these values an amount of energy must be supplied which is proportional to 
$\epsilon_g$.
This energy can transport only
through the interactions with other cells,
therefore
%hence one can conclude that
\be
\icip
\left(
r %|\textbf{x} - \textbf{x}' |
\right) \propto \epsilon_g
\left(
r %|\textbf{x} - \textbf{x}' |
\right)
,
\ee
where $r = |\textbf{x} - \textbf{x}' |$ is a distance between centers of mass of cells.
We introduce the proportionality factor $U_0$,
assume it to be constant in the leading-order approximation,
%(thus it becomes a free parameter of the theory, which can be determined, \textit{e.g.}, empirically),
and obtain the cells' two-body interaction potential function:
\be\lb{e:intcel}
\icip (r) =
\frac{U_0}{\ell} (r - \ello) \, 
%\text{e}^{-(r/\ell)^2}
\exp{\left[-(r/\ell)^2\right]}
+
{\cal O} \left(r-\ell \right)
,
\ee
where 
%$r = |\textbf{x} - \textbf{x}' |$.
% is a distance between centers of mass of cells.
%up to the terms of order $$.
${\cal O}$-terms 
%are assumed to be small. These terms 
can be discarded unless
interactions deform cells so strongly that 
their interior structure 
should be taken into consideration.
%can no longer be neglected.
The potential \eqref{e:intcel}
can be further used in many-body simulations of the Korteweg-type materials including magmas in the ``liquid-dissolved gas'' phase, as discussed in the conclusion.\\

\scn{Foam phase}{s:phg}
If temperature
$T < T_0$
%the fluid's temperature changes so that $(T-T_0)/\eta$ turns negative
%which corresponds to negative $b$'s, 
then
$b$ turns negative, and
the potential density $\lanp$ acquires a conventional Mexican-hat shape,
with local degenerate minima at $|\Psi_e | = \sqrt{\rho_0}$, cf. dotted and dash-dotted curves
in Fig. \ref{f:fpotgau}. 
Therefore, topologically nontrivial solitons must exist which interpolate between the local
minima and maximum at $\Psi = 0$. 

\begin{figure}[htbt]
\begin{center}\epsfig{figure=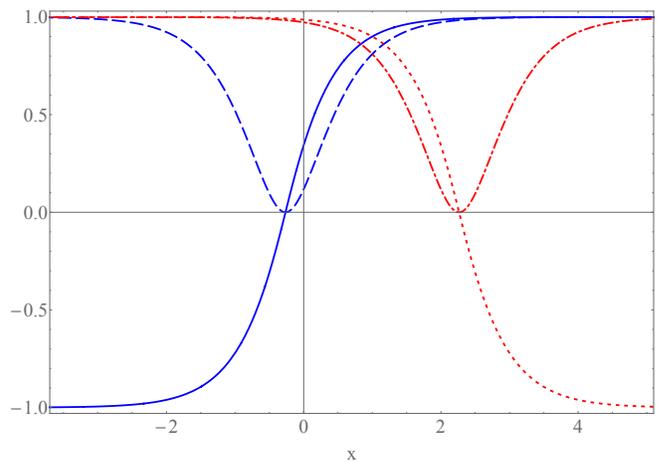,width=  1.0\columnwidth}\end{center}
\caption{
Profiles of kink and antikink solitons $\psi_\pm$ (in units of $\sqrt{\varrho_0}$, solid and dotted curves, 
respectively) 
and their linear densities $\varrho_\pm$ (in units of $\varrho_0$, dashed and dash-dotted curves), versus the Cartesian coordinate $x$ (in units of  $\ell$). For the computations
we used the dimensionless form of \eq \eqref{e:esol} and the boundary condition 
$\psi(+\infty) = 
%- \varrho_0^{-1/2} \psi(-\infty) = 
\pm \sqrt{\varrho_0}$.
%, for the following values of $b$ (in units of inverse time): $1$ (solid curve), $1/2$ (dashed curve), $-1/2$ (dotted curve), $-1$ (dash-dotted curve).
%Two vertical lines represent a condition $|\Psi| \leqslant |\Psi_\text{cut}| < \infty$ which occurs due to the fluid wavefunction's normalization.
%, where a finite cut-off $|\Psi_\text{cut}| > |\Psi_e |$.
%a well with infinite walls, so that the forbidden region starts at $|\Psi| \geqslant |\Psi_\text{cut}| > |\Psi_e |$, where .  
}
\label{f:kink}
\end{figure}

To find these solitons explicitly, let us look for solutions of \eq \eqref{e:o} in a form
of the product
\be
\Psi (\textbf{x}, t) = \prod\limits_{j=1}^\dm \psi_j (x_j, t)
,
\ee
where each of $\psi$'s being normalized separately:
\be\lb{e:normsep}
\int_{X_j} |\psi_j|^2 \drm x_j  = 
\int_{X_j} \varrho_j \, \drm x_j = 
M^{1/\dm}
%> 0
,
\ee
where $\varrho_j = |\psi_j|^2$ and $X_j$ 
are, respectively, linear density and extent of the fluid along the $j$th coordinate.
Then,
due to its separability in Cartesian coordinates,
\eq \eqref{e:o} decomposes into $\dm$ identical 1D equations
of the form:
\be\lb{e:osep}
i \partial_t \psi_j
=
\left[-\frac{\cf}{2} \partial_{x_j x_j}^2
+ |b| 
 \ln\left(|\psi_j|^{2}/\varrho_0 \right)
\right]\psi_j
,
\ee
where $\varrho_0 = \rho_0^{1/\dm}$, and $\partial_{x_j}$ is a spatial derivative with respect to $j$th coordinate.
Therefore, in this case we deal with only one 1D differential equation,
so the index $j$ can be omitted for brevity.
In view of a Galilean invariance,
we resort to a static case $\psi = \psi (x)$
and write this equation in the form
\be
%\partial^2_{xx} 
\frac{\cf}{2} \Der{^2\,\psi}{x^2} = 
\frac{\drm\,
\bar\lanp (|\psi|^2) 
}{\drm \psi} 
, \ee
where 
$\bar\lanp (f) = 
|b| f
\left[
\ln{(f/\varrho_0)} -1
\right]
+ \bar\lanp_0
,
$
$ \bar\lanp_0 = |b| \varrho_0$,
%prime is a derivative with respect to $x$,
and $x$ refers to any of the coordinates $x_1,...,x_\dm$.
According to the standard approach \cite{Rajaraman:1982is}, 
%we can decrease an order of this differential equation, therefore
our solitons 
must saturate the Bogomolny-Prasad-Sommerfield (BPS) bound,
which means that they must be solutions of the first-order differential equation
\be\lb{e:esol}
%\partial^2_{xx} 
%\frac{1}{2} [
\Der{\psi}{x}
%]^2 
= \pm\sqrt{2 \tilde U (\psi)}
,
\ee
where $\tilde U (\psi) = (2/\cf)\bar\lanp (|\psi|^2) $ is a soliton particle potential,
and a sign $\pm$ refers to soliton and anti-soliton solutions.
While the analytical solution for this case is unknown, numerical study reveals
the existence of topological solitons of a kink type, see Fig. \ref{f:kink}.
These solutions have a nonzero topological charge,
\be
Q = \varrho_0^{-1/2}[\psi(+\infty) - \psi(-\infty)]
, \ee
which enhances their stability against decay into a trivial state $\psi = 0$.
Therefore, all nonsingular finite-energy solutions of \eq \eqref{e:esol}
can be cast into four
topological sectors
%, as discussed in Sec. 4 of Ref. 
\cite{gg15}.
Two of these sectors have a nonzero topological charge which ensures stability
of corresponding BPS solitons.

Since the density of each of kink solitons grows from the center of mass outwards, 
cf. dashed and dash-dotted curves in Fig. \ref{f:kink},
these solitons are viewed
as models of bubbles with a characteristic size $\ell$.
While in a single-soliton system these solutions would extend for the whole $x$-axis, in the real fluid
the
kinks would match with antikinks 
%(and \textit{vice versa}) 
at distances of order $\ell$.
%, and \textit{vice versa}. 
Therefore, Korteweg-type magma in this phase tends to form a foam, thus starting 
a process of releasing the previously dissolved gas.\\

\scn{Conclusions}{s:con}
%
%In this \art we 
We have used statistical mechanics arguments
to derive a flow equation for Korteweg-type fluid,
which contains logarithmic nonlinearity making
it similar to wave equations used in the theory of dense Bose-Einstein condensates.
In the classical hydrodynamics context, this equation defines a diffuse interface model
with spontaneous symmetry breaking which causes phase transitions.
It allows multiple (eigen)solutions whose set forms a Hilbert space of different states,
in which the Korteweg fluid can be found.
Therefore, as temperature changes,
fragmentation and nucleation of inhomogeneities occur in erupting magma,
followed
by a transition between the ``magma-dissolved gas'' fluid phase 
and the magmatic foam which flows up the conduit.
We have studied this spontaneous symmetry breaking phenomenon and
related phase structure, by both analytical and numerical methods.

For the cellular phase, we recover a well-known solitary wave solution with a Gaussian density profile,
which describes 
localized cell-like density inhomogeneities in magma.
For such cellular structures
an effective interaction potential and many-body Hamiltonian are derived, which can be used for statistical simulations
of magma and other Korteweg-type materials.
The derived many-body interaction potential \eqref{e:intcel} between Gaussian-shaped density 
inhomogeneities in the liquid phase
of Korteweg fluid suggests a direction for further statistical mechanics' studies of cellular
structures in magmas, where these inhomogeneities
are effectively regarded as point particles with nonsingular interactions.

Considering the foam phase, we demonstrate existence
of the 
topological solitons of a kink type which  
describe bubbles or other inhomogeneities with density increasing from the center of mass outwards, up to 
distances of order $\ell$.
We show that these solitons are saturating BPS bound and belong to a topological 
sector distinct from the one containing the trivial solution.
Thus,
their stability is ensured by the conservation of a topological charge.
The foam-like structure arises in the fluid due to
the
matching between kinks and antikinks 
at distances of order $\ell$;
this structure facilitates release of any gas if it was previously dissolved in the fluid.

Furthermore, 
%a many-body interaction potential between bubble-like structures in the foam phase is still pending its derivation.
%Finally, 
studies of the logarithmic model's Hilbert space and corresponding eigenfunctions
and eigenvalues will shed light upon transitions in 
Korteweg fluids in general and appropriate kinds of magma in particular.
%In this regard, the statistical density operator approach, somewhat similar to the one used in the Maxwell-\schrod~analogy \cite{z16}, will be worth of study as well.

\bc
***
\ec
\begin{acknowledgments}
Proofreading of the
manuscript by P. Stannard is greatly appreciated.
% as well.
This work is based on the research supported by the National Research Foundation of South Africa 
under Grants Nos.
95965 
%98083 
and 98892.
%This work is based on the research supported wholly / in part by the National Research Foundation of South Africa (Grant Numbers xxx, yyy, and zzz)

\end{acknowledgments}

%\appendix

\end{document}